\documentclass[12pt]{article}
\usepackage{amsfonts}
\usepackage[square,sort&compress, numbers, comma]{natbib}
\usepackage{amsmath,amssymb,array,graphicx}
\usepackage{graphicx}
\usepackage[dvips]{color}
\usepackage{subfigure}
\usepackage{multirow}
\usepackage{natbib}
\usepackage{authblk}

\newcommand{\beq}{\begin{equation}}
\newcommand{\eeq}{\end{equation}}
\newcommand{\bey}{\begin{eqnarray}}
\newcommand{\eey}{\end{eqnarray}}
\newcommand{\nn}{\nonumber}

\def\ds{\displaystyle}

\def\p{{\partial}}

\usepackage{setspace}
\usepackage{sectsty}
\sectionfont{\large}

\begin{document}
\newpage
\title{ An Integral Equation Approach for the Valuation of\\ Finite-maturity margin-call Stock Loans}

\author[a]{Minh-Quan Nguyen \thanks{e-mail 
quan.m.nguyen@vn.ey.com}}
\author[b]{Nhat-Tan Le \thanks{e-mail 
tanle@fulbright.edu.vn}}
\author[c,d]{Khuong Nguyen-An\thanks{e-mail: nakhuong@hcmut.edu.vn}}
\author[e,f]{Duc-Thi Luu \thanks{e-mail 
duc-thi.luu@devinci.fr}}
\affil[a]{\small {Ernst \& Young Vietnam Limited, Bitexco Financial Tower, 20F, 2 Hai Trieu, Ben Nghe Ward, District 1, Ho Chi Minh City, Viet Nam}}
\affil[b]{\small {Undergraduate Faculty, Fulbright University Vietnam, 105 Ton Dat Tien, Tan Phu Ward, District 7, Ho Chi Minh City, Vietnam}}
\affil[c]{Faculty of Computer Science and Engineering,
Ho Chi Minh City University of Technology (HCMUT), 268 Ly Thuong Kiet Street, District 10, Ho Chi Minh City, Vietnam}
\affil[d]{Vietnam National University Ho Chi Minh City, Linh Trung Ward, Thu Duc City, Ho Chi Minh City, Vietnam}
\affil[e]{\small {Léonard de Vinci Pôle Universitaire, De Vinci Research Center, 92916 Paris La Défense, France}}
\affil[f]{\small {Department of Economics, University of Kiel, 24118 Kiel, Germany}}
%
	\date{\today}
	\maketitle
\begin{abstract}

This paper examines the pricing issue of margin-call stock loans with finite maturities under the Black-Scholes-Merton framework. In particular, using a Fourier Sine transform method, we reduce the partial differential equation governing the price of a margin-call stock loan into an ordinary differential equation, the solution of which can be easily found (in the Fourier Sine space) and analytically inverted into the original space. As a result, we obtain an integral representation of the value of the stock loan in terms of the unknown optimal exit prices, which are, in turn, governed by a Volterra integral equation. We thus can break the pricing problem of margin-call stock loans into two steps: 1) finding the optimal exit prices by solving numerically the governing Volterra integral equation and 2) calculating the values of margin-call stock loans based on the obtained optimal exit prices. By validating and comparing with other available numerical methods, we show that our proposed numerical scheme offers a reliable and efficient way to calculate the service fee of a margin-call stock loan contract, track the contract value over time, and compute the level of stock price above which it is optimal to exit the contract. The effects of the margin-call feature on the loan contract are also examined and quantified.


 
\end{abstract}

\smallskip\noindent

\vspace{-0.8cm}
\smallskip\noindent
\vspace{-0.6cm}


\vspace{-0.2cm}
\section{Introduction}
\vspace{-0.3cm}


Over the last few decades, critical events in the financial markets and economic fluctuations in different credit cycles have implied a vital role of collateral in modern economies \cite{Gai_et_al_2006, Gorton_Ordonez_2014, Becard_Gauthier_2022}. Much of the lending is, indeed, secured by different forms of collateral, ranging from mortgaged properties, physical assets, and a pool (securitization) of other loans to stocks \cite{Geanakoplos_Zame_2014, Dou_et_al_2019}. On top of that, collateral can be re-pledged from party to party such that the same one can serve multiple lending contracts, thus allowing agents to overcome the scarcity of collateral and increase the liquidity in the whole financial system \cite {BOTTAZZI_et_al_2012, LUU_et_al_2021, MAURIN2022}.  Therefore, estimating the true value of the collateral and understanding how their values can vary over time are meaningful to credit risk management and crucial for safeguarding the system's liquidity and stability.

This paper focuses on the pricing problem of a stock loan with finite maturity. In principle, such a loan contract allows a client (borrower) to obtain a loan amount from a financial institution (lender) by using the borrower's stock 
as collateral. The borrower may retrieve the share at any time before or on the loan maturity date by simply repaying the accumulated debt (the initial loan amount plus a predetermined interest). A margin-call feature may be added to the loan contract to provide more protection to the lenders in exchange for a lower service fee to the borrower. More specifically, if the stock price falls on the accumulated debt, the borrower of a margin-call stock loan has to pay back a pre-determined percentage of the debt. Once the borrower pays back the required amount, called the margin-call payment, the loan contract continues on as a non-recourse stock loan (i.e., there is no margin-call payment if the stock price falls again on the accumulated debt). For the sake of simplicity, we assume here that only one margin call is allowed, and the lender will collect the dividend before redemption. The value of a stock loan with multiple margin calls can be recursively computed based on the value of the corresponding stock loan with one margin call.

Indeed, the stock loan market has been expanding worldwide \cite{ Pang_Wang_2020, McWalter_Ritchken_2022}.
Stock (share) pledging, margin call, and their effects on the governance and decisions of firms have also received remarkable attention in corporate finance research. For example, \citet {Dou_et_al_2019} examined insiders' use of stock shares as collateral for their personal bank loans. Their findings indicate that the margin calls triggered by significant drops in stock prices may increase the crash risk of pledging firms.  Furthermore, the stock pledging may curb a firm's willingness to take risks because margin calls, once they occur, may cause insiders to suffer personal liquidity shocks or to forego private benefits associated with controlling the firm.
In another study, \citet{Chan_et_al_2018} found that controlling shareholders may initiate share repurchases to fend off potential margin calls associated with pledged stocks to maintain their control rights. In response to such behaviour, investors will discount the potential benefits of the repurchases.  In more recent work, \citet{Huang_et_al_2022} studied the consequences of stock pledge restrictions on investment in China and found that, from a policymakers' perspective, restrictions on stock pledge can help to improve the investment efficiency, particularly for private firms, firms with weak financing constraints and firms with high risk-taking.  Evidence on the influence of the use of shares as collateral and margin call pressures on the performance and governance of firms can be found in several other studies such as in \cite{Chan_et_al_2018, Ni_et_al_2022, Pang_Wang_2020, Liu_et_al_2022}.

Nevertheless, few studies have attempted to model how the value of the stock loan can change over time in the presence of fluctuations in the underlying stock's price and the possibility of margin calls.
\citet{XiaZhou} are, perhaps, pioneers in solving this problem.
They used a purely probabilistic approach to price a non-recourse stock loan as a perpetual American call option with a time-dependent strike price. Several authors extended the study to stock loans with
non-standard features, such as capped
stock loan \cite{LIU20103548}, and stock loan with automatic termination clause, cap, and margin \cite{LiangWJ10}. Stock models were also extended from geometric Brownian motion to more complicated ones. Examples include stochastic volatility models \cite{WONG2012337},  jumps or exponential phase-type Levy
models \cite{WONG2013275, Cai_Sun_2014}, regime-switching models \cite{Zhang98}, and models with stochastic or floating interest rates \cite{Chen_et_al_2015, Wang_Chen_2020}.

All the above studies assumed that the stock loan has infinite maturity.
In practice, most stock loan contracts have finite maturities instead. 
Until now, only a few studies have attempted to price a margin-call stock loan with finite maturity. Just as recently, \citet{LU201612} used the Laplace transform method to price such a stock loan. More specifically, the authors used the Laplace transform to reduce the partial differential equation (PDE) governing the price of a margin-call stock loan into an ordinary differential equation (ODE), the solution of which can be numerically found in the Laplace space. As a result, a pricing formula for the stock loan value was obtained in the Laplace space.  The Stehfest method, a numerical Laplace inversion, was then applied to obtain the value in the original space.


Contributing to a richer understanding of the topic, 
 in our present work, we propose to use the continuous Fourier Sine transform (FST) instead of the Laplace transform method used in \cite{LU201612} to reduce the PDE governing the price of a margin-call stock loan into an ODE. The reason for this approach is that the obtained ODE in the Fourier Sine space can be easily found and analytically inverted into the original space. There is no numerical step required to convert values in the Fourier Sine space to those in the original space. As a result, compared with the numerical scheme proposed in \cite{LU201612}, less computational time is required to compute the stock loan value thanks to our newly derived pricing formula in the original space. This is evidenced through our numerical validation and discussion in Section \ref{comparison}. In addition, the effects of the margin-call feature on the stock loan contract value and the contract service fee are examined and quantified.


The remainder of our paper is organized as follows. Section 2 establishes the governing PDE system of a margin-call stock loan with finite maturity. The analytical solution procedure is presented in Section 3, while the numerical implementation and validation are discussed in Section 4. Concluding remarks are given in Section 5.

\section{The governing PDE system}
\vspace{-0.3cm}

We shall now establish the PDE system governing the value of a margin-call stock loan with finite maturity. To begin with, let us assume that the stock loan contract has the following specifications:
\begin{itemize}
	\item At time $t=0$, a stock owner uses his/her stock share as collateral to borrow an amount $E$ of money from a lender, with a constant interest rate $\eta$ compounded continuously for a period $[0, T]$, whereas the lender charges an amount $c$ ( $0\le c \le E$) for the service fee. 
 The borrower, thus, gets the amount of $(E - c)$ from the lender.
	\item At any time $t>0$, if the stock price falls on the accumulated loan amount $Ee^{\eta t}$, a margin call is issued. If this is the case, the loan contract is then temporally terminated until the borrower pays back a predetermined percentage, denoted by $\Delta$, of the accumulated loan amount to the lender. Once the borrower fulfils the obligation, the loan contract becomes a non-recourse one, since the lender now can only take nothing else than the stock (collateral) if the borrower defaults on the loan.  
	\item The borrower has a right, but not an obligation, to regain the stock at any time $t\in(0, T]$ by paying the accumulated loan to the bank.
	\item The lender will receive continuous stock dividends with rate $\delta$ until the borrower regains the stock.
	\item After expiry, the lender will own the stocks if the borrower has not paid fully the accumulated loan.
\end{itemize}
In this valuation problem, at each time $t>0$, we determine both the value of the stock loan contract and the optimal exit price (the level of stock price above which it is better to exit the loan contract than hold it). 
In general, we know that the price of a margin-call stock loan may depend on the current time $t$, the current stock price $S_t$, the principal loan $E$, the stock volatility, the risk-free interest rate, the loan interest rate, and the expiry time  \cite{Hull2018}. In this paper, we use the Black-Scholes-Merton framework discussed in \cite{Black_Scholes_1973, Hull2018}. In particular, we assume that the asset price process $(S_t)_{t\ge 0}$ with a continuous constant dividend yield $\delta$ follows geometric Brownian motion given by
\beq dS_t=(\mu-\delta)S_tdt+\sigma S_t dZ_t,\label{1.2.1}\eeq
where $(Z_t)_{t\ge 0}$ is a standard Brownian motion, $\sigma$ is a constant stock volatility, $\mu$ is a constant drift rate.
The stock loan value can then be expressed as a two-variable function $V(S,\tau)$ of stock price $S$ and time to maturity $\tau$ (i.e, $\tau=T-t$), where $T$ is the contract expiry date.  
Under the Black-Scholes-Merton framework, when the stock loan has not been exited, its value $V(S,\tau)$ should satisfy the classical  equation:
\beq \ds\frac{\p V}{\p \tau}=\frac{\sigma^2 S^2}{2}\frac{\p^2 V}{\p S^2}+(r-\delta)S\frac{\p V}{\p S}-rV,\label{1.2.2}\eeq
where $r$ is a constant risk-free interest rate.

Note that the borrower has a right to regain the stock at any $\tau$ by paying the accumulated loan amount $a(\tau)$, where $a(\tau)= Ee^{\eta (T-\tau)}$, to the lender. The difference between the stock price and the accumulated debt, $S-a(\tau)$, is called the ``exit payoff". Usually, the borrower considers exiting the loan contract only when the exit payoff is positive. To avoid arbitrage opportunities, for each $\tau$, the stock loan value should be always positive and at least equal to the exit payoff. In particular, if the stock loan value is strictly greater than the exit payoff, it is better to hold the loan contract than exit it. On the other hand, if the stock loan value is exactly equal to the exit payoff, the borrower may consider exiting the contract. In this case, holding the contract would not bring any extra benefit than exiting the contract while the loan interest still must be paid.

Hence, for each $\tau$, there exists a critical price $S_f(\tau)$, called ``optimal exit price", which is the smallest value of stock price such that $V(S,\tau)= S-a(\tau), ~\forall S\ge S_f(\tau)$. 
Under the assumption of continuity of the asset price path and dividend yield, it can be shown that these optimal exit prices form a continuous curve using arguments similar to those presented in \cite{kwok2008mathematical}. This curve, called the ``optimal exit boundary", divides the pricing domain into two parts:  1) the below part, the ``holding region", where holding the stock loan is better than retrieving back the stock; and 2) the above part, the ``exiting region", where the stock loan can be exited optimally. 

The pricing formula for the stock loan is thus known in the exiting region. In particular, at the optimal exit curve, we have  \begin{equation}\label{1.2.3}
V(S_f(\tau),\tau) = S_f(\tau) - a(\tau), ~~~~\forall~ 0\leq \tau\leq T.
\end{equation}
In order to determine the unknown optimal exit price $S_f(\tau)$, the following smooth pasting condition is required \cite{lu2015semi}:   

\begin{equation}
\frac{\p V}{\p S}(S_f(\tau),\tau)= 1~~~~\forall~ 0\leq \tau\leq T.\label{1.2.4}
\end{equation}
The stock loan should be exited at expiry only if the stock price exceeds the accumulated loan. Mathematically, at expiry ($\tau=0$), the value of the stock loan can be expressed as: \beq V(S,0)  =\max(S-Ee^{\eta T}, 0).\label{1.2.5}\eeq

%
%
%
%
%

Also, by definition, if the stock price falls on the accumulated loan amount $a(\tau)$ for the first time at $\tau$, the borrower must immediately pay back a predetermined percentage, say $\Delta>0$, of the accumulated loan amount to the lender. 
Once the margin-call payment $\Delta a(\tau)$ is paid, the stock loan contract becomes a non-recourse one with the corresponding loan amount $(1 - \Delta)a(\tau)$. 
Mathematically, this can be expressed by the following equality \cite{LU201612}:
\beq \ds V(a(\tau),\tau)=V_{st}(a(\tau),\tau; (1 - \Delta)a(\tau))- \Delta a(\tau).\label{1.2.6}\eeq
Here $V_{st}(a(\tau),\tau; (1 - \Delta)a(\tau))$ denotes the value of the corresponding non-recourse stock loan at $\tau$, with the loan amount equals $(1 - \Delta)a(\tau)$ and the stock prices equal $a(\tau)$.

\begin{figure}[http]
    \centering
\includegraphics[width=0.7\linewidth]{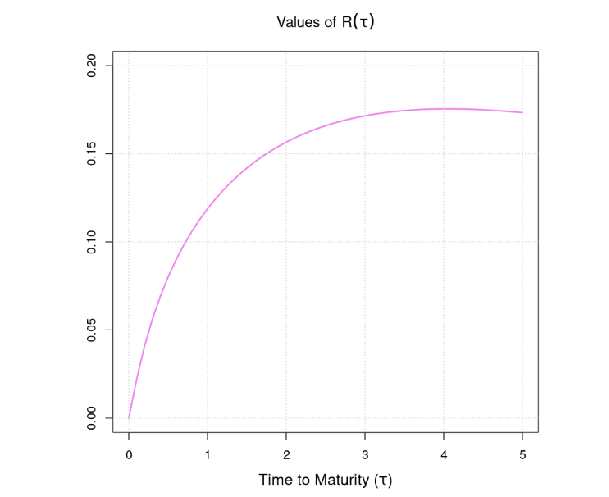}
    \caption{Plot of $R(\tau)$ associated with parameters: $E = 1$; $\eta = 0.1$; $r = 0.06$; $\delta = 0.03$; $\sigma = 0.4$; $T = 5$.}
    \label{rebate}
\end{figure}

All taken together, equations (\ref{1.2.2})-(\ref{1.2.6}) constitute the PDE system governing the value of the stock loan in the holding region. For convenience, they are summarized as
\begin{equation}
\hspace{-.4cm}\mathcal{A}\begin{cases} \label{I}
\ds\frac{\p V}{\p \tau}=\frac{\sigma^2 S^2}{2}\frac{\p^2 V}{\p S^2}+(r-\delta)S\frac{\p V}{\p S}-rV,
~~\text{for} ~~0<\tau \le T, ~ a(\tau) < S < S_f(\tau)\\
\vspace{.15cm} V(S,0)  =\max(S-a(0),0),~~\text{for} ~~ a(0)\le  S \le S_f(0)\\
\vspace{.15cm} \ds V(S_f(\tau),\tau) = S_f(\tau) - a(\tau),~~\text{for} ~~0 \le \tau \le T\\
\vspace{.15cm} \ds \frac{\p V}{\p S}(S_f(\tau),\tau)= 1,\text{for} ~~0\le\tau \le T \\
\vspace{.15cm} \ds V(a(\tau),\tau)=R(\tau), ~~\text{for} ~~0\le \tau \le T,
\end{cases}
\end{equation}
where  $$R(\tau)= V_{st}(a(\tau),\tau; (1 - \Delta)a(\tau))- \Delta a(\tau).$$
Figure \ref{rebate} illustrates how the values of $R(\tau)$ change with respect to $\tau$ for a given set of parameters.
The values of $R(\tau)$ can be obtained by using an integral equation approach as presented by \citet{LE230319}.
 Note that, at $\tau=0$, we have: $$ R(0)=V_{st}(a(0),0; (1 - \Delta)a(0))- \Delta a(0)= a(0)-(1 - \Delta)a(0)-\Delta a(0)=0.$$

The system (\ref{I}) can be solved by using the continuous Fourier Sine transform (FST) proposed by \citet{tanbarrier2015}, which allows us to find $S_f (\tau)$,  the optimal exit price, and $V(S, \tau)$,  the stock loan value. The method will be discussed in detail in the next section. 

 \section{Our analytical solution procedure}
 \label{solution}
 To derive an expression for $V(S, \tau)$ amendable to computation, the PDE system (\ref{I}) is first reduced to a dimensionless heat equation in a finite time-dependent domain.
Then by using FST, the resulting heat equation can be further reduced to an initial value ODE, the solution of which can be easily obtained in the Fourier sine space and analytically inverted back to the real space coordinate.

We now first introduce the dimensionless variables:
\begin{eqnarray*} \ds && S = a(\tau) e^x,~\tau=\frac{2}{\sigma^2}\ell, ~S_f(\tau) = a(\tau) e^{x_f(\ell)},~V(S,\tau) = E e^{\alpha x+\beta l} C(x,\ell),~ R(\tau) = E U(\ell).\end{eqnarray*}
System (\ref{I}) now becomes a dimensionless system, which includes a standard heat equation together with the following corresponding initial and boundary conditions
\begin{equation}
\mathcal{B}\begin{cases} \label{II}
\ds \frac{\p C}{\p \ell}(x,\ell)= \frac{\p^2 C}{\p x^2}(x,\ell), \\
\vspace{.15cm} \ds C(x,0)  =f(x),\\
\vspace{.15cm} \ds C(x_f(\ell),\ell)=  g_1(x_f(\ell),\ell), \\
\vspace{.15cm} \ds \frac{\p C}{\p x}(x_f(\ell),\ell)= g_2(x_f(\ell),\ell), \\
\vspace{.15cm} \ds C(0,\ell)=e^{\beta \ell}U(\ell),
\end{cases}
\end{equation}
where $f, g_1, g_2$ are functions defined as
\begin{eqnarray}\label{f_g}
&&f(x) = \max(e^{(1-\alpha)x+\eta T}-e^{-\alpha x+\eta T},0),\nn\\
&&g_1(x,y) = e^{(1-\alpha)x -\beta y+\eta(T - \frac{2y}{\sigma^2})}-e^{-\alpha x -\beta y+\eta(T - \frac{2y}{\sigma^2})},\nn\\
&&g_2(x,y) =  (1-\alpha)e^{(1-\alpha)x -\beta y+\eta(T - \frac{2y}{\sigma^2})}+\alpha e^{-\alpha x -\beta y+\eta(T - \frac{2y}{\sigma^2})},\nn\\
&&\ds\gamma=\frac{2r}{\sigma^2},~ q=\frac{2\delta}{\sigma^2}, ~ k = \gamma - q - \frac{2\eta}{\sigma^2}-1,~\alpha = - \frac{k}{2}, ~\beta = -\alpha^2 -\gamma,
\end{eqnarray}
and $\mathcal{B}$ is defined on $\ds \ell\in [0,T\frac{\sigma^2}{2}],~x\in[0, x_f(\ell)]$.


Our next step is to apply FST to the system (\ref{II}).
For readers' convenience, we recall the definition of the FST of a function here. More specifically, the FST of $C(x,\ell)$ with respect to $x$, denoted by $\mathcal{F}_s\{C(x,\ell)\}$, is defined as
$$\mathcal{F}_s\big\{C(x,\ell)\big\} \equiv U(\omega,l) = \int_0^{\infty}C(x,\ell)\sin(\omega x)dx,$$
with the corresponding inversion
$$\mathcal{F}_s^{-1}\big\{U(\omega,l)\big\} = \frac{2}{\pi}\int_0^{\infty}U(\omega,l)\sin(\omega x)d\omega.$$
 As we will use the continuous Fourier cosine transform (FCT) in our solution procedure later, we also recall here the definition of FCT and its inversion as
$$
\mathcal{F}_c\left\{C(x,\ell)\right\}\equiv U(\omega,l)=\int_0^\infty C(x,\ell)\cos(\omega x)dx, \quad \mathcal{F}^{-1}_c\left\{U(\omega,l)\right\} = \frac{2}{\pi}\int_0^{\infty}U(\omega,l)\cos(\omega x)d\omega,
$$
respectively.

To apply FST, the $x$-domain of (\ref{II}), which is a finite domain, needs to be extended to a semi-infinite domain first by expressing the PDE as
\begin{equation} 
\label{n301} H(x_f(\ell)- x)\frac{\p C}{\p \ell}(x,\ell)= H(x_f(\ell)- x)\frac{\p^2 C}{\p x^2}(x,\ell)
\end{equation}
where $H(x)$ is the Heaviside function, defined as
\begin{equation}
H(x) = \begin{cases}
\ds 1, ~~~~~~~{\rm{if}}~~~~x > 0,\\
\ds 1/2, ~~~~~~{\rm{if}}~~x = 0,\\
\ds 0, ~~~~~~~~{\rm{if}}~~~x <0.
\end{cases}
\end{equation}
The reason for the appearance of the factor of $1/2$
at the point of discontinuity is explained in \cite{chiarella2004survey}. The initial and boundary conditions remain unchanged.

As a result of applying the FST with respect to $x$, the PDE (\ref{n301}) can be reduced to the following linear first-order ODE in the Fourier sine space
 \begin{eqnarray*}\frac{\p \widehat{C}}{\p \ell}(\omega,\ell)+\omega^2\widehat{C}(\omega,\ell)= G(\omega,\ell) \end{eqnarray*}
where
\vspace{-1cm}\bey G(\omega,\ell)&=& \sin \big(\omega x_f(\ell)\big)\frac{\p C}{\p x}\big(x_f(\ell),\ell\big) - \omega \cos\big(\omega x_f(\ell)\big)C\big(x_f(\ell),\ell\big)+\omega C(0,\ell)\nn\\
&&+ x_f'(\ell)C\big(x_f(\ell),\ell\big)\sin\big(\omega x_f(\ell)\big),\nn \eey
with initial condition $\ds\widehat{C}(\omega,0) = \int_0^{x_f(\ell)}C(x,0)\sin(\omega x) dx$. The solution of this initial-value ODE can be easily solved as
 \begin{equation}\label{i1}
\widehat{C}(\omega,\ell)= \widehat{C}(\omega,0)e^{-\omega^2\ell} +\int_0^{\ell}e^{-\omega^2(\ell - \xi)}G(\omega,\xi)d\xi.
 \end{equation}
As $\widehat{C}(\omega,\ell)$ denotes the Fourier sine transform of $H\big(x_f(\ell)- x\big)C(x,\ell)$, from (\ref{i1}), we can now express the solution of the system (\ref{II}) as follows
 \begin{equation}\label{i2}
H\big(x_f(\ell)- x\big)C(x,\ell)= \mathcal{F}^{-1}_s\bigg\{\widehat{C}(\omega,\ell)\bigg\} = \mathcal{F}^{-1}_s\bigg\{\widehat{C}(\omega,0)e^{-\omega^2\ell}\bigg\}  +\mathcal{F}^{-1}_s\bigg\{\int_0^{\ell}e^{-\omega^2(\ell - \xi)}G(\omega,\xi)d\xi\bigg\},
 \end{equation}
where $\mathcal{F}^{-1}_s$ is the notation for the corresponding inversion of the  Fourier transform.
 
 Furthermore, as $S = a(\tau)e^x, ~S_f(\tau) = a(\tau)e^{x_f(\ell)}$, and $~V(S,\tau) = Ee^{\alpha x+\beta \ell} C(x,\ell)$, by multiplying $Ee^{\alpha x+ \beta \ell}$ to both sides of (\ref{i2}), we can express the solution of the system (\ref{I}) as follows
\begin{equation}\label{i3}
H\big(\ln S_f(\tau) - \ln S\big) V(S,\tau) = Ee^{\alpha x+ \beta l}\mathcal{F}^{-1}_s\bigg\{\widehat{C}(\omega,0)e^{-\omega^2l}\bigg\} +Ee^{\alpha x+ \beta \ell}\mathcal{F}^{-1}_s\bigg\{\int_0^{\ell}e^{-\omega^2(\ell - \xi)}G(\omega,\xi)d\xi\bigg\}.
\end{equation}

After a tedious calculation of the inversion of FST, similar to those presented in \cite{tanbarrier2015}, the first term on the right-hand side of (\ref{i3}) can be expressed as
\begin{equation}\label{first_term}
 Ee^{\alpha x+ \beta \ell}\mathcal{F}^{-1}_s\{\widehat{C}(\omega,0)e^{-\omega^2\ell}\}
=M(S,\tau,a(0))- M\big(S,\tau,S_f(0)\big)
\end{equation}
where the function $M$ is given by
\begin{equation}\label{M_def}
M(x,y,z) = M_1(x,y,z)-\left(\frac{x}{a(y)}\right)^{\lambda} M_1\left(\frac{(a(y))^2}{x},y,z\right).
\end{equation}
Here, $M_1$ is defined by
\begin{equation*}
M_1(x,y, z) = xe^{-\delta y}N(d_1(x,y,z))-ze^{-ry}N(d_2(x,y,z)),
\label{eq:def_M_1}
\end{equation*}
with
\begin{eqnarray}\label{gen_def}
&& N(x) = \frac{1}{\sqrt{2\pi}}\int_{-\infty}^x e^{-u^2}du, ~~d_1(x, y, z) = \frac{\ln x - \ln z + (r - \delta + \sigma^2/2)y}{\sigma\sqrt{y}},\nn \\
 && d_2(x, y, z) = \frac{\ln x - \ln z + (r - \delta - \sigma^2/2)y}{\sigma\sqrt{y}},~ \lambda = 2\alpha, ~a(y)= Ee^{\eta (T-y)}.
 \end{eqnarray}

The second term in the right-hand side of (\ref{i3}) can be expressed as
\begin{eqnarray}\label{second_term}
&&Ee^{\alpha x+ \beta \ell}\mathcal{F}^{-1}_s\bigg\{\int_0^{l}e^{-\omega^2(\ell - \xi)}G(\omega,\xi)d\xi\bigg\} \nn\\
&=& - (S - a(\tau)). 1_{S = S_f(\tau)}(S) + M\big(S,\tau,S_f(0)\big)+\int_0^{\tau}Q\big(S,\tau,u,S_f(u)\big)du,
\end{eqnarray}
where $M$ is defined as in \eqref{M_def} and
\begin{equation}\label{Q_def}
Q(x,y,z,w) = Q_1(x,y,z,w)- \left(\frac{x}{a(y)}\right)^{\lambda}Q_1\left(\frac{(a(y))^2}{x},y,z,w\right)+ \left(\frac{x}{a(y)}\right)^{\frac{\lambda}{2}} K(x,y,z).
\end{equation}
Here, $Q_1$ and $K$ are defined by
\begin{eqnarray*}
&&Q_1(x,y,z,w)= x \delta e^{-\delta (y-z)}N(d_1(x,y-z,w))- a(y)(r-\eta)e^{-(r-\eta)(y-z)}N(d_2(x,y-z,w)),\nn\\
&&K(x,y,z)=\frac{\ln x -\ln a(y)  }{\sigma\sqrt{2\pi}\sqrt{(y-z)^3}}e^{-\frac{(\ln x - \ln a(y))^2}{2\sigma^2(y-z)}+\beta\frac{\sigma^2}{2}(y-z)} R(z),
\end{eqnarray*}
with
\begin{eqnarray*} 1_{S = S_f(\tau)}(S) = \begin{cases}
\ds \frac{1}{2}, ~~~~~~~{\rm{if}}~~~~S = S_f(\tau),\\
\ds 0, ~~~~~~~~{\rm{if}}~~~ S \ne S_f(\tau),
\end{cases}
 \end{eqnarray*}
 and $N, d_1, d_2, \lambda, a(y)$ are defined as in \eqref{gen_def}.



From the formulas \eqref{first_term} and \eqref{second_term}, we can now obtain an integral representation of the price of a margin-call stock loan.
More specifically, the stock loan value can be expressed in terms of its optimal exit prices as follows
\begin{eqnarray}\label{main_result}
&&V(S,\tau) 
=M\big(S,\tau,a(0)\big)+\int_0^{\tau}Q\big(S,\tau,u,S_f(u)\big)du, \forall S < S_f(\tau)
\end{eqnarray}
with $M$ and $Q$ are functions defined in \eqref{M_def} and \eqref{Q_def}. Here, the optimal exit prices satisfy the following integral equation
\begin{equation}\label{IE}
S_f(\tau)-a(\tau)=M\big(S_f(\tau),\tau,a(0)\big)+\int_0^{\tau}Q\big(S_f(\tau),\tau,u,S_f(u)\big)du.
\end{equation}

To highlight the effect of the margin-call feature on the pricing of a stock loan contract, we now compare the integral representations of stock loans with and without margin calls.
The integral representation for the value of a non-recourse stock loan derived by \citet{LE230319} as follows:
\begin{eqnarray}\label{non_recourse}
&&V(S,\tau) 
=M_1\big(S,\tau,a(0)\big)+\int_0^{\tau}Q_1\big(S,\tau,u,S_f(u)\big)du, \forall S < S_f(\tau)
\end{eqnarray}
By comparing the two formulas (\ref{main_result}) and (\ref{non_recourse}), we can deduce that the three terms:
\begin{eqnarray*}
&&\displaystyle\left(\frac{S}{a(\tau)}\right)^{\lambda} M_1\left(\frac{(a(\tau))^2}{S},\tau,a(0)\right),\\ &&\displaystyle\int_0^{\tau}\left(\frac{S}{a(\tau)}\right)^{\lambda}Q_1\left(\frac{(a(\tau))^2}{S},\tau,u,S_f(u)\right)du,\\
 &&\displaystyle\int_0^{\tau}\left(\frac{S}{a(\tau)}\right)^{\frac{\lambda}{2}} K(S,\tau,u)du
\end{eqnarray*}
 appearing in (\ref{main_result}) represents the effect of the margin-call feature on the stock loan value. These terms also add more computational expense to the pricing problem.

 \section{Numerical implementation, validation and discussions}

In this section, we solve numerically the integral equation (\ref{IE}) to approximate the optimal exit prices by 
using a numerical procedure
discussed by \citet{kallast2003pricing} and \citet{tanbarrier2015}. The obtained results can then be used to calculate the value of the stock loan via formulas (\ref{main_result}). Our numerical results are validated and compared with other available numerical methods. The interesting effects of the margin-call feature on the stock loan contract are also discussed.

\subsection{Numerical procedure}
We shall now briefly describe the numerical procedure to approximate the optimal exit price. To begin with, 
let $\Pi$ be a uniform partition of the time to maturity interval $[0, T]$ representing the life of the stock loan such that
\begin{equation*}
\Pi: 0 = x_1< x_2 < \ldots <x_n < x_{n+1} = T, \text{with the same incremental step } h = \frac{T}{n}.
\end{equation*}
We aim to find the optimal exit price at each of these $(n+1)$ discrete points. To this end, we first approximate the integral equation (\ref{IE}) at each discrete point $x_i$ by an algebraic equation. 
More specifically, the first algebraic equation comes from the fact that $$	\ds S_f(0^{+}) =\max\left(Ee^{\eta T}, \frac{r-\eta}{\delta}Ee^{\eta T}\right).$$ This can be proved by taking the limit $t\to T$ (i.e. $\tau \to 0$) of both sides of equation \eqref{IE}, in a way similar to that presented in \cite{chiarella2004survey}. 
It is clear that $S_f(0^{+})$ should be at least equal to the accumulated debt, as the borrower will not regain the stock if its value is less than the accumulated debt. In practice, for compensating the inherent risk in lending activity, the lending interest rate $\eta$ is usually greater than the risk-free interest rate $r$ so that we usually have the equality $S_f(0^{+})=S_f(0)=Ee^{\eta T}.$ In this case, the optimal exit boundary is thus continuous at expiry.

We now also consider extreme cases where the risk-free interest rate is greater than the sum of the lending rate and dividend rate, i.e., $r>\eta+\delta$. We then obtain $S_f(0^{+})=\dfrac{r-\eta}{\delta}Ee^{\eta T}$. 
This financially means that as the time approaches expiry, the borrower should exit the stock loan only if the stock price $S$ is large enough so that the sum of the obtained dividend $S\delta dt$ (in a very short time period $dt$) and the paid loan interest $\eta Ee^{\eta T}dt$ is at least equal to the earned saving interest $rEe^{\eta T}dt$. This is due to the fact that when exiting the loan contract near expiry, the obtained benefit for the borrower is to receive the continuous dividend from the retrieved stock and does not have to pay the loan interest any more, while
the corresponding cost is to repay the accumulated loan $Ee^{\eta T}$ and thus can not earn the saving interest from that amount.
 In these extreme cases, $S_f(0^{+})> Ee^{\eta T}=S_f(0)$ so that the optimal exit boundary is discontinuous at expiry. 

To solve for $(n+1)$ unknown variables $\{S_f(x_i)\}_{i =1}^{n+1}$, we need $n$ more algebraic equations, which can be derived  by substituting $\tau = x_j, j \geq 2$ into the integral equation (\ref{IE}) as follows
\begin{equation} \label{ap1}
S_f(x_j)-a(x_j)=M\left(S_f(x_j),x_j,Ee^{\eta T}\right)+\ds\int_0^{x_j}Q\left(S_f(x_j),x_j,u,S_f(u)\right)du,
\end{equation}

where $\ds\int_0^{x_j}Q\left(S_f(x_j),x_j,u,S_f(u)\right)du$ can be further decomposed to two terms: 
$$\displaystyle\int_0^{x_j}\left[Q_1\big(S_f(x_j),x_j,u,S_f(u)\big)+\left(\frac{S_f(x_j)}{a(x_j)}\right)^{\lambda}Q_1\left(\frac{(a(x_j))^2}{S},\tau,u,S_f(u)\right)\right]du$$ and
$$\displaystyle\int_0^{x_j}\left(\frac{S_f(x_j)}{a(x_j)}\right)^{\frac{\lambda}{2}} K(S,x_j,u)du.$$

The first integral term can be approximately by the simple Trapezoidal rule as the integrand is a smooth function. 
The integrand $K$, however, in the second integral term exhibits a singularity at $u=x_j$, making it challenging to approximate the integral accurately.
To overcome this issue, we can employ a variable transformation
$$v=K_2(x,y)\left(\sqrt{\frac{y}{y-z}}-1\right),K_2(x,y)=\frac{\ln x-\ln a(y)}{\sigma\sqrt{2y}},$$
and convert the second term into an integral over a semi-infinite domain
\begin{align*}
\int_0^yK(x,y,z)dz
= \frac{2}{\sqrt{\pi}}\int_0^{\infty}e^{-v}K_3(x,y,v)dv
\end{align*}
where 
$$K_3(x,y,v)=\left(\frac{x}{a(y)}\right)^{\alpha}\cdot R\left(y-\frac{y\cdot K_2(x,y)^2}{(v+K_2(x,y))^2}\right)\cdot e^{\left(\frac{(\ln x-\ln a(y))^2\beta}{4(v+K_2(x,y))^2}-(v+K_2(x,y))^2+v\right)}.$$

This transformed integral can then be approximated using Gauss-Laguerre quadrature rules, which provide more accurate and stable results. 
As a result, at each $x_j$, (\ref{ap1}) can be reduced to an algebraic equation. We now have a system of nonlinear algebraic equations, which can be solved recursively by using the Newton-Raphson iteration procedure to obtain the optimal exit price at each discrete point in time. 
It should be noted that the value of the embedded non-recourse stock loan is an important input to calculate the value of the margin-call stock loan. This input can be computed by using the Fourier transform method, as presented in \cite{LE230319}.


\subsection{Numerical validation and discussion}
 \label{comparison}

In this section, we will validate our proposed approach by showing the agreement between the results of our integral equation (IE) method, the standard binomial tree (BT) method and the Laplace transform method proposed in \cite{lu2015semi, LU201612}. For the BT method, 10,000 time steps are employed to obtain reliable results. On the other hand, for our IE method, only 50 time steps are utilized to obtain similar results.\footnote{Our numerical results are obtained  using Rstudio, version: 2024.04.2+764 on a PC with the following details: Processor 12th Gen Intel(R) Core(TM) i7-1255U 1.70 GHz, installed Ram 16 GB
RAM, Windows 11 enterprise service pack, 64-bit operating system.}

We begin by validating our numerical scheme for a non-recourse stock loan. 
Table \ref{tab1} below showcases the calculations of the optimal exit prices at inception. From this table, it is clear that our numerical results agree well with those calculated by using a binomial tree method and a Laplace transform method presented in \cite{lu2015semi}. In addition, the average running time of our method is significantly faster than the other two methods. This is because our numerical scheme does not require any numerical step for inverting back values in the Fourier Sine space to the original space.
\begin{table}[htbp]
	\centering
	\caption{ Optimal exit price at inception of a non-recourse stock loan with parameters:
		$E=0.7$,  $\eta=0.1$, $r=0.06$, $\sigma =0.4$, $\delta=0.03$.}
	\vspace{2ex}
 \label{tab1}
	\begin{tabular}{|c|c|c|c|c|c|c|} 
		\hline
	 ~ &  \multicolumn{2}{c|}{IE}
		& \multicolumn{2}{c|}{\citet{lu2015semi}}
		& \multicolumn{2}{c|}{Binomial ($N=10{,}000$)} \\
		\hline
		$T\text{(year)}$ &  $S_f$  & times
		& $S_f$  & times
		& $S_f$  & times
		\\ \hline
	
		1    &1.168       &0.10              &1.197      &  0.21     & 1.163-1.168    &57.2
		
		\\ \hline
		5    &1.529       &0.13              &1.527      &  1.02     & 1.524-1.538    &57.6
		
		\\ \hline
		20    & 1.843       &0.17              &1.825      & 3.77     & 1.839-1.872    &57.5
		
		\\ \hline
	
	\end{tabular}
\end{table}

To further validate our IE method for margin-call stock loans, in Table \ref{tab3}, we compare the obtained numerical results for stock loan values derived from IE and BT methods with a given parameter set. The two methods agree well on the stock loan values at inception, with at least one decimal digit, for various values of stock prices and initial loan amount.

\begin{table}[htbp]
    \centering
    \setlength{\tabcolsep}{15pt} 
    \caption{Comparison of margin-call stock loans at inception derived from IE and BT methods for different initial loan amounts $E$ and stock prices $S$, and fixed parameters: $\eta=0.05,r=0.1,\delta=0.05,\sigma=0.2,\Delta=0.1,T=1.$}
    \vspace{2ex}
    \label{tab3}
    \begin{tabular}{|c|c|c|c|c|c|c|}
        \hline
        \textbf{S} & \multicolumn{2}{|c|}{$E=80$} & \multicolumn{2}{|c|}{$E=85$} & \multicolumn{2}{|c|}{$E=90$} \\
        \hline
        & \textbf{IE} & \textbf{BT} & \textbf{IE} & \textbf{BT} & \textbf{IE} & \textbf{BT} \\
        \hline
        95  & 15.291 & 15.289 & 10.848 & 10.846 & 6.738 & 6.735 \\
        100 & 20.062 & 20.062 & 15.373 & 15.372 & 10.970 & 10.968 \\
        105 & 25.000 & 25.000 & 20.106 & 20.105 & 15.462 & 15.461 \\
        110 & 30.000 & 30.000 & 25.001 & 25.003 & 20.159 & 20.158 \\
        \hline
    \end{tabular}
\end{table}


\subsection{The effect of the margin-call feature on stock loan contract}
Introducing the margin-call feature in a stock loan contract reduces the risk to the lender by requiring the borrower to pay the margin-call payment if the margin-call condition is triggered. This feature is expected to decrease the contract's value as the borrower's rights are significantly affected. The computation of the optimal exit boundary and the contract value for a margin-call stock loan is more expensive than that of a non-recourse stock loan due to the presence of the extra mathematical terms associated with the margin-call feature, as highlighted in section 3. We will quantify these effects in this section.

\begin{table}[htbp]
	\centering
	\caption{Optimal exit prices at inception, $S_f(T)$, and at expiry, $S_f(0)$, of margin-call stock loans with fixed parameters:
		$E=0.7$,  $\eta=0.1$, $r=0.06$, $\sigma =0.4$, $\delta=0.03$, $\Delta=0.1$.}
	\vspace{2ex}
 \label{tab2}
	\begin{tabular}{ccccccc}
		\hline
		$\textbf{T\text{(year)}}$ & & $S_f(T)$  & &$S_f(0)$& & \textbf{Computing time (seconds)}
		\\ \hline
		
		$1$     & &$ 1.043$   &    &$0.774 $ & &$1.44$           
		
		\\ 
	$5$    & &$1.358$      & &$1.154 $ & &$1.47$  
		\\ 
	$10$    & &$1.509$      & &$1.902 $ & &$1.56$            
		\\ \hline
		
	\end{tabular}\label{t1}
\end{table}

 As illustrated in Table \ref{tab2}, with the same parameters, using the IE method, the average computation time of the optimal exit boundary for a margin-call stock loan is about 1.5 seconds, which is ten times slower than that of a corresponding non-recourse.

We now also examine the effect of different values of the margin-call payment proportion $\Delta$ on the optimal exit boundary.
\begin{figure}[htbp]
	\centering
	\includegraphics[width=0.6\linewidth]{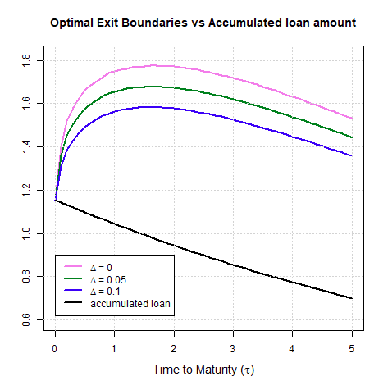}
	\caption{Optimal exit boundaries and accumulated loan amount at different values of $\Delta$, and fixed parameters $E=0.7$,  $\eta=0.1$, $r=0.06$, $\sigma =0.4$, $\delta=0.03$, $T=5$, $\Delta=0.1$}
	\label{deltachange}
\end{figure}
  As we can see from Figure \ref{deltachange}, all three optimal exit boundaries exhibit a hump shape behaviour. As time elapses (i.e., $\tau$ decreases), the accumulated loan principal increases, and the optimal exit prices also naturally tend to increase so that the obtained exit payoff is high enough to cover the increasing debt. As long as the time to expiry is still large enough, there may exist a big gap between the optimal exit price and the accumulated loan amount, as there is enough time to expect a high level of stock price. When time gets closer to expiry ($\tau\to 0$), this gap naturally becomes smaller as there is less room for stock to vary. Our numerical results also show that the optimal exit boundary with a higher value of $\Delta$ stays below the one with a lower value of $\Delta$. This phenomenon is financially intuitive since a higher level of $\Delta$ implies a larger margin-call payment, and hence, the optimal exit boundary should be lower to increase the chance of exiting the contract before the margin-call event occurs.

\begin{figure}[htbp]
	\centering
	\includegraphics[width=0.7\linewidth]{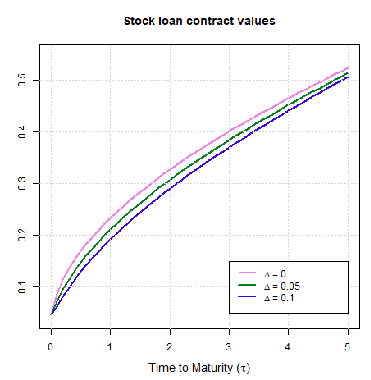}
	\caption{Variation in stock loan contract value over $\tau$ for different values of $\Delta$, at a fixed stock price $S=1.2$, and fixed parameters $E=0.7$, $\eta=0.1$, $r=0.06$, $\sigma =0.4$, $\delta=0.03$, $T=5$, $\Delta=0.1$}
	\label{deltachange_value}
\end{figure}

\begin{figure}[htbp]
	\centering
	\includegraphics[width=0.7\linewidth]{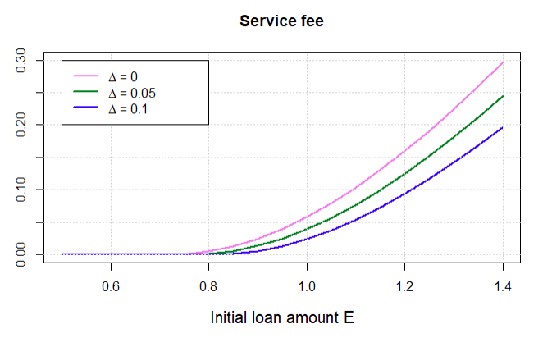}
	\caption{Variation in stock loan contracts' service fees over different initial loan amount $E$, different values of $\Delta$, and fixed parameters $\eta=0.1$, $r=0.06$, $\sigma =0.4$, $\delta=0.03$, $T=5$, $\Delta=0.1$}
	\label{deltachange_service_fee}
\end{figure}
A higher value of $\Delta$ provides more protection to the lender and thus should result in a lower price for entering a margin-call loan contract. To illustrate this fact, in Figure \ref{deltachange_value}, we examine the stock loan contract values at different $\tau$ for a fixed stock price value. The Figure clearly shows that the contract with a lower value of $\Delta$ is more expensive than that of a higher value of $\Delta$.

Consequently, the service fee for a stock loan contract with a higher margin-call payment should be less than that of a lower margin-call payment. Note that this service $c$ is computed as $c= V_0-S_0+E$, where $V_0$ and $S_0$ are the contract value and stock price at inception, respectively. Figure \ref{deltachange_service_fee} also illustrates the fact that the higher the initial loan amount, the higher the service fee, reflecting a higher risk to lenders.

\section{Conclusion}
\vspace{-0.2cm}

This paper presents a semi-analytical approach to determining the value of a margin-call stock loan contract with finite maturity, its optimal exit boundary, and the associated service fee. Under the Black-Scholes-Merton framework, we can establish a PDE system governing the movement of the stock loan value with both fixed and free boundary conditions. By using the Fourier Sine transform method, proposed in \cite{tanbarrier2015}, we can express the value of the stock loan value in terms of its optimal exit prices, which in turn is governed by an integral equation.  We demonstrate that the optimal exit prices form a continuous curve called the ``optimal exit boundary". Such a boundary separates the pricing domain into two distinct areas: the ``holding region", where it is more advantageous to keep the stock loan, and the ``exiting region", where exiting the stock loan is optimal. In addition,  we are able to quantify mathematically the effect of the margin-call feature on the value of the stock loan contract, the optimal exit boundary, and the service fee. We also propose an efficient numerical scheme to produce desired numerical results quickly and reliably.

From our present work, several research directions can be proposed. For example, it would be interesting to consider different stochastic processes governing the stock price used as collateral in the loan contract. In addition, from the initial framework introduced in our paper,  further analysis of the relationship between changes in the value of the collateral and the estimation of loss-given default could provide valuable insights for credit risk management. Moreover, we believe that integrating our framework with equity dynamics, especially considering the impacts of dilution and share buybacks \cite{McWalter_Ritchken_2022, Backwell_et_al_2022}, might open interesting avenues for future research. Exploring the application of the valuation method for finite-maturity stock loans, as developed in our current work, to pricing issues of non-financial products like recallable tickets or the ``optimal
launch time" for highly fashionable items in business \cite{Chiu_et_al_2017} could also be a fruitful research topic.

\section*{Acknowledgments}

We are very grateful to participants in the various conferences, workshops, and seminars where earlier versions of this paper were presented. These include the European Consortium for Mathematics in Industry (Poland, June 2023), the academic seminar at the Fulbright University of Vietnam (Spring, 2023), the visiting research trip at Quy Nhon University (June 2024), and the second annual meeting of the International Society for the Advancement of Financial Economics (Thailand, July 2024). All remaining errors are, of course, our own.

\section*{Disclosure of interest}
All authors declare that they have no conflicts of interest.



\vspace{1cm}

\appendix
\makeatletter
\def\@seccntformat#1{\csname Pref@#1\endcsname \csname the#1\endcsname\quad}
\def\Pref@section{Appendix~}
\makeatother
\vspace{-0.5cm}

\bibliographystyle{apalike}

\linespread{0.1}
{\small \bibliography{review3}}

\end{document}